\documentstyle[aps,twocolumn]{revtex}

\begin{document}
\draft
\wideabs{
\title{Magnetic properties of superconducting
multifilamentary tapes in perpendicular field. I: Model and vertical
stacks}
\author{Alvaro Sanchez$^{\rm a}$, Carles
Navau$^{\rm a,b}$, and Enric Pardo$^{\rm a}$, }
\address{$^{\rm a}$ Grup d'Electromagnetisme,
Departament de F\'\i sica, Universitat Aut\`onoma Barcelona \\
08193 Bellaterra (Barcelona), Catalonia, Spain\\
$^{\rm b}$ Escola Universit\`aria Salesiana de Sarri\`a, Passeig Sant
Joan Bosco 74, 08017 Barcelona, Catalonia, Spain}
\maketitle
\begin{abstract}
Current and field profiles, and magnetization and AC losses are
calculated for an array of infinitely long superconducting tapes
arranged vertically in a perpendicularly applied magnetic field.
Calculations are based on the critical state model. The finite
thickness of the tapes and the effects
of demagnetizing fields are considered. The influence of the magnetic
coupling of the filaments in the magnetic properties of the arrays
are studied. The general model can be applied to an arbitrary
arrangement of tapes as long as there is reflection symmetry with
respect to the vertical central plane. 
\end{abstract}
}

\section{Introduction}

The magnetic behavior of hard superconductors depends not only on
their intrinsic properties, such as the critical-current density
$J_c$, but on their geometry. For example, the magnetization and ac
losses for a superconducting thin film are very different depending
on whether the magnetic field is applied perpendicular or parallel to
the thin
dimension.
Recent advances in the modeling of magnetic properties of
superconductors (SCs) have allowed to understand their behavior in a
large variety of geometries. The most common framework used to
address this problem has been the critical state-model 
\cite{bean}, that assumes that currents circulating in the SCs
flow with a constant density $J_c$, later extended to currents
depending only upon the local magnetic field
$H_{\rm i}$ \cite{kim}. The original model was solved in the parallel
geometry, that is for applied fields $H_{\rm a}$ along an infinite
dimension for slabs and cylinders \cite{bean,kim,chengoldfarb},
because
in these geometries the problem of the demagnetizing effects was
not present. A further step
was presented when the CSM was extended to the case of very thin
strips \cite{thinstrips,zeldov,mcdonald} and disks
\cite{Mikheenko,Zhu,ClemSanchez}, for which
important demagnetizing fields were involved. More recently, the more
general case of critical state in samples with finite thickness, such
as strips \cite{thickstrips} and cylinders
\cite{thickcylinders,doyle,prb1} was
solved by numerical models.

However, there is a particular arrangement that has not been solved
so far in spite of its practical importance. It is the case of a set
of superconductors with finite thickness arranged in an array form.
The importance of this case
is not only academic but practical, since this geometry is often a
good system for modeling the filamentary structure of actual
superconducting
tapes. The recent advances in the technology of superconducting
Ag/Bi2223 multifilamentary tapes increase the importance of having a
theory describing the magnetic properties of such arrays.

Although the problem of studying the current distribution
and magnetization in array of superconducting tapes has not been
systematically
solved, there have been significant works offering partial solutions.
Fabbricatore et al \cite{Fabbricatore} presented a comprehensive
analysis of the Meissner state in arrays of strip lines arranged
vertically ($z$ stack of strips), horizontally ($x$ arrays) and in
the form of a matrix ($xz$ array) and compared their results with
actual measurements on multifilamentary tapes. Their numerical
procedure, however, was not adequate to study the more general case
of bulk current penetration. Mawatari \cite{Mawatari} studied not
only the Meissner state but also the critical state for the
case of an infinite set of periodically arranged
superconducting-strip lines, in the limit that the strip lines were
infinitely thin. Mawatari and Clem \cite{MawatariClem} studied the
penetration of magnetic flux into current-carrying (infinitely thin)
strips lines with slits in the absence of applied magnetic field. 
All the existing models assume either arrays of infinitely thin
strips in the critical state or
arrays of strips with finite thickness but only in the Meissner
state. 
The only exception we know is the recent work by Tebano et al
\cite{Tebano} in which preliminary results on the current penetration
and magnetization were calculated for some realistic arrays based on
the
procedure developed by Brandt \cite{thickstrips}.

A key issue in the study of superconducting tapes is
how currents circulate within filaments. There are two important
cases
concerning this point, depending on whether current in each filament
is restricted to go
and return through the same filament or if instead there is no such
restriction. In
principle, both cases
have practical and theoretical interest, as we discuss in the second
paper of this series \cite{tapesb}. The magnetic
response and AC losses will strongly depend on this factor so it is
imperative to study both cases separately and compare them.

In this series of papers we study the current and field penetrations,
magnetization and AC losses of arrays of superconducting strips of
finite thickness. In the first paper we present the model and its
application to the case of an array of a finite number of infinitely
long strips of
finite thickness arranged vertically ($z$ stack of finite strips)
with a perpendicular applied field.
This geometry is studied first because it is independent of the
connection type, since
because of the system symmetry, current always go and return through
each strip.
In the second paper we will study the cases of horizontal ($x$)
arrays
and matrices ($xz$ array), for which different behaviors arise
depending on the connections. In all cases, we will concentrate our
study in tapes composed of strips with high aspect ratio since
this is the case most often met in practice, although our model is
applicable to arrays of strips with arbitrary thickness. However,
this is different from
the approximation of 
considering infinitely thin strips as in \cite{Mawatari}, since we
take into account the different current penetration across the
superconductor thickness.

The present paper is structured as follows. In section II we
introduce the calculation model. Current and field profiles are
calculated and discussed in Section III.  The results of
magnetization and ac
losses are discussed in Section V and VI, respectively. In Section
VII we present the main conclusions of this work. In two appendices
we present analytical expressions for both the full penetration field
of the arrays and the inductances used in the model calculations.

\section{Model}

The model we present here is suitable for any superconductor geometry
with translational symmetry along the $y$ axis and mirror symmetry to
the vertical plane. However, we will focus on $z$-stacks, $x$-arrays,
and $xz$-matrices made up of identical strips infinitely long in the
$y$ direction. The horizontal and vertical strips dimensions are $2a$
and $2b$, respectively. We consider a uniform applied field $H_{\rm
a}$ in the $z$ direction. Results for $z$-stacks are discussed in the
present paper, whereas those for $x$-arrays and $xz$-matrices are
presented in the second paper of this series \cite{tapesb}.

Our numerical model is based on minimizing the magnetic energy of the
current distribution after each applied field variation. The model
assumes there is no equilibrium
magnetization in the superconductor and, in the present series of
papers, we further assume there is no field dependence of
$J_c$ for simplicity.  The energy
and flux
minimization in the critical state was previously discussed
by Badia \cite{Badia} and Chaddah et al \cite{Chaddah,Chaddah2}. The
details of the general numerical model can be found in
\cite{Alvaro1,prb1,prb2,sst}. The 
approach has been successfully applied to describe the experimental
features observed in the initial magnetization slope \cite{Alvaro1}
as well
as the whole magnetization loop and levitation force \cite{prb1,prb2}
of superconducting cylinders. 
We now outline the main characteristics
of the model.

For cylindrical geometry \cite{Alvaro1,prb1,prb2}, the
superconducting
region was divided into a certain number of elements of the shape of
rings with rectangular cross-section, which formed circular closed
circuits. For a given current distribution in the cylinder, the
energy variation of setting a new current in a certain element was
calculated. The minimum energy
variation method consists in, given an applied field
$H_{\rm a}$, looking for the circuit which lowers the most the
magnetic energy
and set there a new
current of magnitude $J_c$, and then repeat the procedure until
setting
any new current does not reduce the energy. The
method as
explained above is used to find the initial magnetization curve
$M_i(H_{\rm a})$. Provided that $J_c$ is field independent, the
reverse
curve can be found using that \cite{ClemSanchez} 
\begin{equation}
\label{reverse1}
M_{\rm rev}(H_{\rm a})=M_i(H_{m})-2M_i((H_{m}-H_{\rm a})/2),
\end{equation}
and the returning curve using 
\begin{equation}
\label{reverse2}
M_{\rm ret}(H_{\rm a})=-M_{\rm rev}(-H_{\rm a}), 
\end{equation}
where $H_m$ is the maximum applied field in the loop.

In the present case, each strip is divided into a set of $2n_x\times
2n_z$ elements with cross-section
$(\Delta x)(\Delta z)$ and an infinite length in $y$ direction, as
shown in Fig. 1. The dimensions of each element are $\Delta
x=a/n_x$
and $\Delta y=b/n_y$. We consider that
the current density is uniform within the elements
and flows through the whole element section and not only through a
linear circuit, as in \cite{Alvaro1,prb1,prb2}.

When the distribution and orientation of the strips in the set are
symmetrical to the $zy$ plane, as is the case for a $z$-stack, in the
presence of a uniform applied field the induced current front will
also be symmetrical to the $zy$ plane. Thus,
we can consider that the pair of elements centered at $(x,z)$ and
$(-x,z)$ form circuits that are closed at infinity. This
grouping in pairs, forming closed circuits, allows for the analytical
calculation of self and
mutual inductances per unit length of the circuits with finite
cross-section (see the Appendix B
for inductances derivations and formulae). 
The pairs, or circuits, are labelled using the subscript $i$ from 1
to
$N=2n_x n_z
n_f$, being $n_f$ the number of strips of the set and $N$ the total
number of elements in the $x\ge 0$ portion of the set of strips.

Once the analytical expressions for the inductances are obtained, the
energy of the $i$ circuit can be calculated as
\begin{equation}
\label{energy}
E_{i}=\sum_{j=1}^N M_{ij}I_{j}I_{i}+2\mu_0 H_{\rm a} x_i I_i,
\end{equation}
where the first term is the energy of the circuit owing to the
presence of the current distribution in the whole
superconducting region, the second term is the energy due to the
uniform applied magnetic field, $M_{ij}$ are the self and
mutual inductances from Eqs. (\ref{Mij})-(\ref{gdef}), and $I_{i}$
and $I_{j}$ are the total current intensity that flows through the
circuits labelled as $i$ and $j$, respectively. Since no internal
field dependence is considered for the critical current,
$|I_i|=(\Delta x)(\Delta z)J_c$. 
The sign of $I_i$ 
is taken as positive when the current of the element at $x\ge 0$ of
the pair follows the positive $y$ axis direction and negative
otherwise.

In the initial magnetization curve, after using the energy
minimization procedure for a given applied field $H_{\rm a}$ to find
the
current profile, we can calculate the magnetization, the total
magnetic field, and the magnetic field lines directly form the
current distribution.

The magnetization, defined as the magnetic moment per unit
volume, has only one non-zero component, $M_z$, which can be
calculated as
\begin{equation}
\label{mag}
M_z={{m_z}\over{4abn_f}}={1\over 4abn_f}\sum_{i=1}^N I_i
\left({2x_i}\right),
\end{equation}
where $m_z$ is the total magnetic moment of the set of strips.

The two nonzero components of the total magnetic flux density $B_z$
and $B_x$ are calculated as the addition of all the closed circuit
contributions, which can be calculated analytically integrating the
Biot-Savart law \cite{Landau}.

The magnetic flux lines are calculated as in \cite{thickstrips}
using that for translational symmetry the level curves of the $y$
component of the vector potential, for the gauge $\nabla \cdot
{\bf A}=0$, can be taken as the magnetic flux lines. The $y$
component of the vector potential results from the contribution
of all the elements pairs $A_{y,i}$, whose analytical expressions are
the Eqs. (\ref{Ayj})-(\ref{fdef}). 

We will consider the
application of an external AC field $H_{\rm a}=H_{m}\cos(\omega t)$.
In this case we can 
calculate the imaginary part of the AC susceptibility, $\chi''$,
defined as
\begin{eqnarray}
\label{chi}
\chi''&=&{2\over \pi H_m}\int_0^\pi d\theta M_{\rm rev} (\theta)
\sin\theta \nonumber\\
&=&{2\over \pi H_m^2}\int_{-H_m}^{H_m} dH_{\rm a} M_{\rm rev} (H_{\rm
a}).
\end{eqnarray}
The
energy loss $W$ per AC cycle of amplitude $H_{\rm a}$ is related to
$\chi''$ by the expression
\cite{spectra,ClemSanchez}
\begin{equation}
\label{losses}
W=\mu_0\pi H_{\rm a}^2\chi''
\end{equation}

\section{Current penetration and field lines}

The most important issue to study in the system of
superconducting strips is the influence of magnetic
coupling. It is well known that the superconductors tend to shield
the magnetic field change not only in their interior but also in the
space between two superconducting regions (as illustrated in the
classical example of a hollow superconducting tube, for which the
whole volume inside the tube and not only the tube walls are
shielded). In our array of superconducting strips one should observe
a related effect, because the current induced in the strips will tend
to shield the effect of the magnetic field in their interior as well
in the space between them. To study this effect we present in Fig. 2
calculations corresponding to a vertical array of three strips with
$b/a=0.1$ and different separations ($h=2a$, $h=0.2a$, and
$h=0.02a$,
respectively). The different profiles correspond to applied fields
$H_{\rm a}/H_{\rm pen}=$0.2, 0.4, 0.6, 0.8, and 1, where $H_{\rm
pen}$ is the field at which the array is fully penetrated by current
(the calculation of such a field is discussed in Appendix 1). The
results show unambiguously the strong influence of magnetic coupling
in the case that the separation is small, as illustrated in the case
for $h=0.02a$ for which the current profiles are almost the same as
if there were no gaps between the strips. On the other hand,
the case $h=2a$ is already an example of a very weak magnetic
coupling that result in a current penetration for each strip almost
as if the two others were not present. 

To further study this point we present in Fig. 3 the field lines
calculated for the three arrays of Fig. 2, for applied fields
$H_{\rm a}/H_{\rm pen}=0.4$. The left images show the total field and
the right ones only the field created by the currents
in the superconductor. It is clear that for $h=0.02a$ and even
$h=0.2a$,
the applied field in the space between superconductors is basically
shielded by them, in contrast with the case of $h=2a$, for which
the magnetic field is modified near each strip but not enough to make
a significant contribution to the other two strips. Another way of
seeing
this effect is illustrated in the calculations of the field created
by currents (right images in each figure). One can observe
that in all cases currents create a basically constant field
in the spaces between strips. However, when the separation is large
the total
field lines created by the current are wrapped around each tape,
whereas when the separation decreases up to $h/a=0.02$ the field
lines are hardly distinguishable from the case of the three tapes
forming a single thicker one.

\section{Magnetization}

In this section we analyze the magnetization of the arrays,
calculated from the currents following Eq. (\ref{mag}). The reverse
and returning curve can be obtained from the initial one using Eqs.
(\ref{reverse1}) and (\ref{reverse2}). 

There are several important properties of the tapes arrays that can
be understood from the magnetization results.
In fig. 4 we plot the calculated magnetization $M$ as function of
the applied field $H_{\rm a}$ for a set of three arrays, each with
semisides ratio $b/a$
of 0.01 and with different separations $h/a=0.02, 0.2$, and 2,
respectively. We also plot the magnetization for a single tape with
$b/a=0.01$, another one with $b/a=0.03$ corresponding to the
case that the three tapes are one on top of each other. 
One can
observe that the general trend is that the saturation magnetization
remains the same for all cases, whereas the initial slope of the
magnetization curve changes. The slope
is largest in absolute value for the case of a
single thin tape ($b/a=0.01$), has intermediate values for the three
arrays (the largest slope is for the array with large separation and
smallest for the
one with the smallest one), and finally the case with $b/a=0.03$,
which, interestingly, is hardly distinguishable from the array with
the smallest separation. 
The reason for such an enhancement of the initial slope is the
demagnetizing effect associated with the large sample aspect ratio,
according to which the thinner the sample the larger the initial
slope \cite{prb1,thickstrips}. In the
arrays for which the separation between the tapes is small, the
magnetic coupling increase,
in agreement with the
discussion in section III, so that the sample is behaving as
having a larger thickness, and therefore, less demagnetizing effects
and less initial slope. In order to study in more detail this effect,
we have included in Fig. 4 the calculated magnetization of a single 
tape with $b/a=0.05$; this tape corresponds to the array with
$h/a=0.02$ but
as if the gaps between the tapes were filled by superconducting
material as well. We can see that this case has             
the smallest slope, and there are large differences with respect
to the case of the array of three tapes with separation $h/a=0.02$.
From these results we can
conclude that, provided that the strips are close enough, the
behavior of a $z$ array is similar to that of a strip of thickness
the sum of the superconducting regions of the $z$-array, and not the
sum of the whole $z$-array including the gaps.

Another interesting feature to study is the effect of the addition of
more strips to the array.
We compare in Fig. 5 the calculated $M(H_{\rm a})$ curves for
arrays with a fix distance ($h=0.2$) and
different number of tapes. We include in the figure the two
analytically known limits of one infinitely thin strip
\cite{thinstrips} and for an
infinite set of strips \cite{Mawatari}. Results show a practical
coincidence between the calculated results for a single tape with
finite although small thickness and the results from the analytical
formula for very thin strips \cite{thinstrips}. With adding more
tapes, the initial slope of the magnetization (defined as magnetic
moment divided by volume, so independent of the superconductor
volume) gets smaller in absolute value. We find that even the
case of 25 strips is significatively different from the Mawatari case
for an infinite
stack, so we can conclude that Mawatari's formula should be valid
only for
a very large number of tapes. In Fig. 6 we show the
magnetization curves $M(H_{\rm a}$) for arrays of tapes separated a
smaller distance ($h/a=2$). It can be observed that now the
differences in the slope are smaller than in the previous case,
because there is less magnetic coupling among the tapes.

\section{AC losses}

In this section we study the imaginary part of the AC susceptibility,
$\chi''$, calculated form the magnetization loops from Eq.
(\ref{chi}). $\chi''$ is directly related to the power losses by Eq.
(\ref{losses}).

We present in Fig. 7 the calculated results for $\chi''$ as function
of the maximum applied AC field $H_{\rm a}$ corresponding to the
magnetization curves of Fig. 4. All curves show a peak at some value
of the
applied field amplitude. It can be seen that the peak corresponding
to the maximum in $\chi''$ (and therefore a change in slope of the AC
losses, since they are proportional to $\chi''$ times $H_{\rm a}^2$
as seen in Eq. (\ref{losses})) is
shifting to higher fields and decreasing in magnitude with decreasing
separation distance. The reason
for that can be obtained from the analogous shifting in
the initial slope of the magnetization shown in Fig. 4. Since the AC
losses are related to the area of the $M(H_{\rm a})$ curve [Eqs.
(\ref{chi}) and (\ref{losses})] and the magnetization saturation is
the same for all cases, the key factor for the losses behavior is the
initial slope governed by the demagnetizing effects as discussed in
Section IV.

Results in Fig. 7 show that the $\chi''(H_{\rm a}$ curve goes as
$H_{\rm a}^n$ for low fields, with $n$ ranging from 1.5 for a strip
with
$b/a=0.01$, corresponding to the high $h/a$ limit for an $x$-array,
to 1.3 for a strip with $b/a=0.03$, being the low $h/a$ limit for
arrays. These values lie between the known limiting values for
infinite slabs ($n=1$) \cite{bean} and thin strips ($n=2$)
\cite{ThStrPRB}.
The corresponding power losses (related to $\chi''$ by Eq.
(\ref{losses}); see inset) go therefore as $H_{\rm a}^n$, with $n$
ranging from 3.5 to 3.3. For high fields, $\chi''$ goes as $H_{\rm
a}^{-1}$ in all cases.

In Figs. 8 and 9 we calculate the results for $\chi''$ as function
of the maximum applied AC field $H_{\rm a}$ with the goal of studying
the effect of adding more tapes in the array. We use the same cases
as in Figs. 5 and 6, corresponding to two different separations in
the arrays ($h/a=2$, and 0.2, respectively). Again, there is a close
relation between the increase in the initial slope of the
magnetization curves and the shifting of the position of the peak in
$\chi''$ to higher fields. It is interesting to comment, however,
that the formulas provided by Mawatari for the case of an infinite
array are not adequate to describe quantitatively the AC losses of a
finite array even if the array consists of up to 25 tapes.

It is clear from the calculations presented in this section 
that in order to reduce the magnetic contribution to the AC losses in
a real tape made of superconducting filaments with a given thickness
one should increase the
coupling of the filaments by decreasing the distance between them
(see Fig. 7). Also, the results in Fig. 8 and 9 show that the AC
losses decrease with increasing the number of filaments while keeping
the distance between them, particularly in the case of a small
separation ($h/a=0.2$, Fig. 9). These results can be understood again
as arising from the effect of demagnetizing fields. When the
superconducting tapes are separated a small distance, the whole tape
is acting basically as a single superconducting tape with a thickness
equal to the sum of the thickness of the tapes, so the demagnetizing
effects are less, the initial
magnetization has a smaller slope (in absolute value) as discussed in
Section V, and the area of the hysteresis loops is less for a given
maximum applied AC field, so that the AC losses are reduced. On the
other hand, when the separation of the tapes gets larger, the
magnetic coupling between them is less so that they act more like
magnetically
decoupled filaments with high aspect ratio and the demagnetizing
effects act more, increasing the AC losses.
It is interesting to notice that when we talk about the separation
$h$ being small one should understand it as small compared with the
horizontal dimension $2a$ and not to the strip thickness $2b$.
Calculations of the same values of $h/a$ for the case $b/a=0.1$
(instead of $b/a=0.01$ as in the results presented above), for
example, yield qualitatively the same effects for the same $h/a$
values.

\section{Conclusions}

We have presented a numerical model for calculating current
penetration and field profiles, and magnetization and AC losses of an
array of superconducting tapes. In this work we have analyzed the
case of an array of vertically arranged superconducting strips. We
have found that the demagnetizing effects have strong influences in
the magnetic response of the tapes and the AC losses appearing when
an AC field is applied. We find that AC losses are reduced when
decreasing the vertical separation between filaments. When the
vertical separation is small as compared with the filaments width,
then the tape is behaving as a single filament with thickness the sum
of the superconducting material.
These
results could be used as guides for designing actual superconducting
tapes. Then, in order to optimize the losses, for filaments with a
fixed aspect ratio, it is preferable
to have a large number of them separated small distances so that
there is a good magnetic coupling
between them, as has been already experimentally found
\cite{GomAC_00}.
The model can be applied to horizontal and matrix arrangements as is
presented in the next paper.

\section*{Acknowledgments}

We thank Fedor G\"om\"ory and Riccardo Tebano for comments.
We thank MCyT project BFM2000-0001, CIRIT project
1999SGR00340, and DURSI from Generalitat de Catalunya for financial
support.

\appendix

\section{Field of full penetration}

In this section we present a simple way to analytically calculate
the field of full penetration of the array, $H_{\rm pen}$, defined as
the minimum applied
field in the initial magnetization curve for which current fills the
whole of the superconducting region. The penetration field can be
calculated in general as minus the field generated by the current
distribution
${\bf H}_J$
in the last induced current point, where ${\bf H}_J=H_J {\bf
\hat{z}}$ \cite{thickstrips,forkl}. So, both the current distribution
at the penetration field
and the last induced current point position ${\bf r}_m$ must be known
to calculate $H_{\rm pen}$.

In the geometry of a set of rectangular cross-section strips ordered
as a $z$-stack we can distinguish two
different cases, depending on whether the number of strips $n_f$ 
is odd or even. Although in both cases the current distribution at
$H_{\rm pen}$ is evident, it is not so for ${\bf r}_m$. The last
induced current point can be found as where the field generated by
the currents is maximum in magnitude, since ${\bf r}_m$ is the last
point where external field is shielded. For the odd number of strips
case, ${\bf r}_m$ is simply the position of the center of the central
strip, although it is not so easy to determine when the number of
strips is even.

For a $z$-stack of an odd number of strips $n_f$ with dimensions $2a$
and $2b$ in the $x$ and $z$ directions separated a distance $h$ the
penetration field is therefore
\begin{eqnarray}
\label{Hpenodd}
H_{\rm pen}(a,b,h,n_f)={J_c\over 2\pi}\Bigg[F_1(0,a,b)+ &&\nonumber\\
\quad 2\sum_{i=1}^{n_f-1\over 2}F_1((2b+h)i,a,b)\Bigg] &&\qquad(n_f \
{\rm odd}),
\end{eqnarray}
where $F_1(u,t,d)$ is defined as
\begin{eqnarray}
\label{F1def}
F_1(u,t,d)&=&2t \bigg\{\arctan{u+d\over t}-\arctan{u-d\over t}+
\nonumber\\
&& (u-d)\ln\left[{{(u-d)^2\over t^2+(u-d)^2}}\right] +\nonumber\\
&& (u+d)\ln\left[{{(u+d)^2\over t^2+(u+d)^2}}\right]\bigg\}.
\end{eqnarray}
Eq. (\ref{Hpenodd}) has been derived using the expression for the
magnetic field created by a
completely penetrated strip with uniform $J_c$ calculated by
direct integration of the Biot-Savart law. The case $n_f=1$
reproduces the known result for the penetration field for a strip
\cite{thickstrips}.

As mentioned above, when a $z$-stack have an even number of strips we
must find the last point where current is induced, at which the self
field $H_J$ is maximum in magnitude. Owing to the symmetry of the
current fronts in the $yz$ plane, this point will be on the $z$ axis.
Thus, only maximization of $H_{J,z}$ along the $z$ axis is needed.
Since the minimization of the field $H_{J,z}$ is different for every
specific value of $n_f$, we only report the result for $n_f=2$, which
is the most important case, concerning the magnetic coupling, for an
even $n_f$. Then, the last induced current point will be at a
position
\begin{eqnarray}
z_m^2&=&{1\over 2}\Big[ -a^2-h\beta+ \nonumber\\
&& \sqrt{(a^2+h\beta)^2+h\beta(2a^2+h^2/2+2\beta^2+h\beta)}  \Big],
\end{eqnarray}
being $z_m$ the $z$ component of ${\bf r}_m$ and $\beta$ defined as
$\beta=2b+h/2$. The penetration field for two strips is
\begin{eqnarray}
H_{\rm pen}(a,b,n_f=2)&=&{J_c\over 2\pi} \big[ F_1(z_m-b-h/2,a,b)+
\nonumber\\
&&F_1(z_m+b+h/2,a,b) \big],
\end{eqnarray}
where the function $F_1(u,t,d)$ is defined in Eq. (\ref{F1def}).

%\appendix

\section{Calculation of inductances}

In this appendix we calculate the self and mutual inductances used in
Eq. (\ref{energy}). These inductances are calculated for closed
circuits of the shape of a pair of identical rectangular infinite
prisms of dimensions $2a'\times 2b'$ carrying uniform current
density.
The prisms are set symmetrically to the $zy$ plane, taking the $y$
axis parallel to the infinite direction. The current of the prism set
in the $x\ge 0$ region is taken positive, while it is taken negative
for the other.

The self and mutual inductances are calculated from the magnetic
energy using the equation \cite{Landau}
\begin{eqnarray}
\label{MijDef}
M_{ij}I_i I_j=W_{ij}& = &\int_{x_i-a'}^{x_i+a'} dx
\int_{z_i-b'}^{z_i+b'}
dz A_{y,j}(x,z)J_i- \nonumber \\
&& \int_{-x_i-a'}^{-x_i+a'} dx \int_{z_i-b'}^{z_i+b'} dz
A_{y,j}(x,z)J_i,
\end{eqnarray}
where $M_{ij}$ is the mutual inductance per unit length of two closed
circuits labelled as $i$ and $j$ respectively, $I_i$ and $I_j$ are
the current intensity flowing through the circuits, $W_{ij}$ is the
magnetic energy of the circuits, $(x_i,z_i)$ is the central position
of the prism in the $x\ge 0$ region of the $i$ circuit, $a'$ and $b'$
the dimensions of the prisms in the $x$ and $z$ directions
respectively, and $A_{y,j}$ is the $y$ component of the vector
potential created by the circuit $j$ taking the gauge $\nabla \cdot
{\bf A}=0$. 

The vector potential $A_{y,j}$ can be calculated by direct
integration leading to
\begin{equation}
\label{Ayj}
A_{y,j}(x,z)={\mu_0 J_j \over
2\pi}\left[{F(x-x_j,z)-F(x+x_j,z)}\right],
\end{equation}
where the function $F(u,v)$ is defined as
\begin{eqnarray}
F(u,v)& = & f(a'-u,b'-v)+f(a'-u,b'+v)+ \nonumber\\
& & f(a'+u,b'-v)+f(a'+u,b'+v),
\end{eqnarray}
defining $f(t,d)$ as
\begin{eqnarray}
\label{fdef}
f(t,d)& = & {1\over 2} \bigg[ td \ln{\left({t^2+d^2}\right)}-3td+
\nonumber \\
&&t^2 \arctan{d\over t}+t^2 \arctan{d\over t} \bigg].
\end{eqnarray}

Taking into account that the current density in the prisms is
uniform, $M_{ij}$ can be deduced integrating Eq. (\ref{MijDef}) using
Eqs. (\ref{Ayj})-(\ref{fdef}), which yield
\begin{eqnarray}
\label{Mij}
M_{ij}&=&{\mu_0\over 16\pi a'^2 b'^2} \big[ G(x_j-x_i,y_j-y_i)
\nonumber\\
&&-G(-x_j-x_i,y_j-y_i)  \big],
\end{eqnarray}
where the function $G(u,v)$ is defined as
\begin{eqnarray}
G(u,v)&=& \sum_{k,l,n,m=1}^2 (-1)^{k+l+n+m}\times \nonumber\\
&&g\big( R(k,n)a'+u,R(l,m)b'+v \big),
\end{eqnarray}
defining $R(i,j)=(-1)^i-(-1)^j$, and the function $g(t,d)$ as
\begin{eqnarray}
\label{gdef}
g(t,d)&=&{24\over 48}t^2d^2-{t^4\over 96}-{dt^3\over 6}\arctan{d\over
t}-{td^3\over 6}\arctan{t\over d}+\nonumber\\
&&{1\over 48}\left({t^4+d^4-6t^2d^2}\right)\ln\left({t^2+d^2}\right).
\end{eqnarray}

\begin{figure}
\caption{Sketch of the array
of superconducting strips.}
\end{figure}

\begin{figure}
\caption{Current profiles for an array of three superconducting tapes
of width $2a$ and height $2b$ separated a distance (a) $h/a$=2, (b)
$h/a$=0.2, and (c) $h/a$=0.02. The profiles correspond to applied
fields $H_{\rm a}/H_{\rm pen}=$0.2, 0.4, 0.6, 0.8, and 1, where
$H_{\rm pen}$ is the penetration field of the array. For the sake of
clarity, the separation, thickness and width of strips are not on
scale.}
\end{figure}

\begin{figure}
\caption{Field lines corresponding to an applied field $H_{\rm
a}/H_{\rm pen}$=0.4, where $H_{\rm pen}$ is the penetration field of
the $z$-stack, for the stacks of Fig. 2. Right and left figures
correspond to the total and self-field magnetic field lines,
respectively. The distances are $d/a=$ 2 (a,b), 0.2 (c,d), and 0.02
(e,f).}
\end{figure}

\begin{figure}
\caption{Initial magnetization as a
function of the applied field for, from left to right: a single tape
with $b/a=0.01$, an array of three tapes with $b/a=0.01 $ separated a
distance $h/a=2$, the same array with a separation distance of
$h/a=0.2$, the same array with a separation distance of $h/a=0.02$, a
single tape with $b/a=0.03$, and a single tape with $b/a=0.05$. In
the inset complete magnetization loops are plotted for the first and
last cases.  }
\end{figure}

\begin{figure}
\caption{Initial magnetization as a
function of the applied field for, from left to right: a single tape
with $b/a=0.01$, the
theoretical expression for thin strips (almost overlapped), a set of
3 tapes with
$b/a=0.01$ each, a set of 5 tapes
with $b/a=0.01$ each, a set of 9 tapes with $b/a=0.01$ each, a set of
25 tapes with $b/a=0.01$ each, and the expression given by Mawatari
(Ref. 15) for an array of infinite number of tapes of dimensions
$b/a=0.01$.
The separation distance between the tapes in all cases is $h/a$=0.2.
}
\end{figure}

\begin{figure}
\caption{Same as Fig. 5, but for a separation $h/a$=2. }
\end{figure}

\begin{figure}
\caption{Imaginary part of the AC susceptibility $\chi''$ as a
function of the amplitude of the AC field $H_{\rm a}$ corresponding
to the curves of Fig. 4, in the same order (the case of a single tape
with $b/a$=0.05 is not plotted in this figure). The corresponding
power losses are shown in the inset.}
\end{figure}

\begin{figure}
\caption{Imaginary part of the AC susceptibility $\chi''$ as a
function of the amplitude of the AC field $H_{\rm a}$ corresponding
to the curves of Fig. 5, in the same order.}
\end{figure}

\begin{figure}
\caption{Imaginary part of the AC susceptibility $\chi''$ as a
function of the amplitude of the AC field $H_{\rm a}$ corresponding
to the curves of Fig. 6, in the same order.}
\end{figure}


\begin{references}

\bibitem{bean}  C. P. Bean, Phys. Rev. Lett. {\bf 8}, 250 (1962).

\bibitem{kim} Y. B. Kim, C. F. Hempstead, and A. R. Strnad, Phys.
Rev. Lett. {\bf 9}, 306 (1962).

\bibitem{chengoldfarb}  D.-X. Chen and R. B. Goldfarb, J. Appl. Phys.
{\bf 66}, 2489 (1989).

\bibitem{thinstrips} E. H. Brandt and M. Indenbom, Phys. Rev. B
{\bf
48}, 12893 (1993).

\bibitem{zeldov} E. Zeldov, J. R. Clem, M. McElfresh, and M. Darwin,
Phys. Rev. B {\bf 49}, 9802 (1994).

\bibitem{mcdonald} J. McDonald and J. R. Clem, Phys. Rev. B {\bf 53},
8643 (1996).

\bibitem{Mikheenko} P.N. Mikheenko and Y.E. Kuzovlev, Physica C {\bf
204},
229 (1993). 

\bibitem{Zhu} J. Zhu, J. Mester, J. Lockhart, and J. Turneaure,
Physica
C {\bf  212}, 216 (1993) .


\bibitem{ClemSanchez}  J. R. Clem and A. Sanchez, Phys. Rev. B {\bf
50}, 9355 (1994).

\bibitem{thickstrips} E. H. Brandt, Phys. Rev. B {\bf 54}, 4246
(1996). 

\bibitem{thickcylinders} E. H. Brandt,
Phys. Rev. B {\bf 58}, 6506 (1998).

\bibitem{doyle} T. B. Doyle, R. Labusch, and R.
A. Doyle, Physica C
{\bf 290}, 148 (1997).

\bibitem{prb1}  A. Sanchez and C. Navau, Phys. Rev. B, {\bf 64},
214506 (2001).



\bibitem{Fabbricatore}  P. Fabbricatore, S. Farinon, S. Innocenti,
and F.
G\"{o}mory, Phys. Rev. B {\bf 61}, 6413 (2000).

\bibitem{Mawatari} Y. Mawatari, Phys. Rev. B {\bf 54}, 13215 (1996).  

\bibitem{MawatariClem} Y. Mawatari and J. R. Clem, Phys. Rev. Lett.
{\bf 86}, 2870 (2001)


\bibitem{Tebano} R. Tebano, F. G\"{o}mory, E.Seiler, F. Stryceck,
Physica C, to be published (2002).

\bibitem{tapesb} E. Pardo, C.Navau, and A. Sanchez, preprint.


\bibitem{Badia}  A. Badia, C. Lopez, and J. L. Giordano, Phys. Rev. B
{\bf 58}, 9440 (1998).

\bibitem{Chaddah} K. V. Bhagwat, S. V. Nair, and P. Chaddah, Physica
C {\bf 227}, 176 (1994)

\bibitem{Chaddah2} P. Chadah, Pranama- J. Phys. {\bf 36}, 353
(1991).

\bibitem{Alvaro1}  F. M. Araujo-Moreira, C. Navau, and A. Sanchez,
Phys.
Rev. B {\bf 61}, 634 (2000)



\bibitem{prb2} C. Navau and A.
Sanchez, Phys. Rev. B,  {\bf 64}, 214507 (2001).

\bibitem{sst}  A. Sanchez and C. Navau, Supercond. Sci. Technol. {\bf
14}, 444 (2001).



\bibitem{Landau} L. D. Landau, E. M. Lifshitz, and L. P.
Pitaevskii, in {\em electrodynamics of Continuous Media} (2nd
edition) (Pergamon Press, New York, 1984).



\bibitem{spectra}  D.-X. Chen and A. Sanchez, J. Appl. Phys. {\bf
70}, 5463
(1991).


\bibitem{forkl} A. Forkl, Phys. Scr. {\bf T49}, 148 (1993). 


\bibitem{GomAC_00} F. G\"om\"ory, J. \v Souc, A. Laudis, P. Kov\'a\v
c, and I. Hu\v sek, Supercond. Sci. Technol. {\bf 13}, 1580 (2000).

\bibitem{ThStrPRB} E. H. Brandt and M. Indebom, Phys. Rev. B {\bf
48}, 12893 (1993).

\end{references}
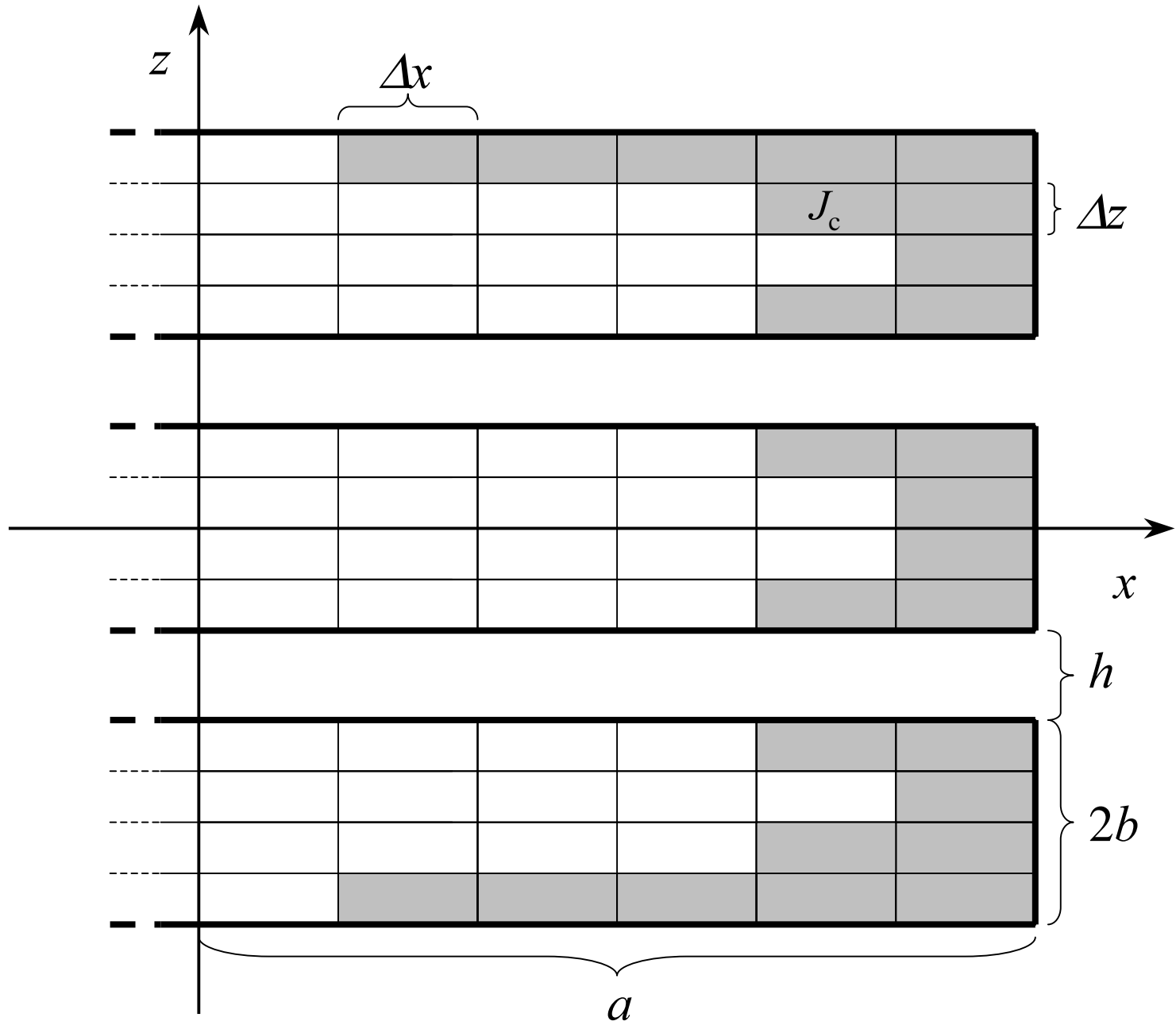
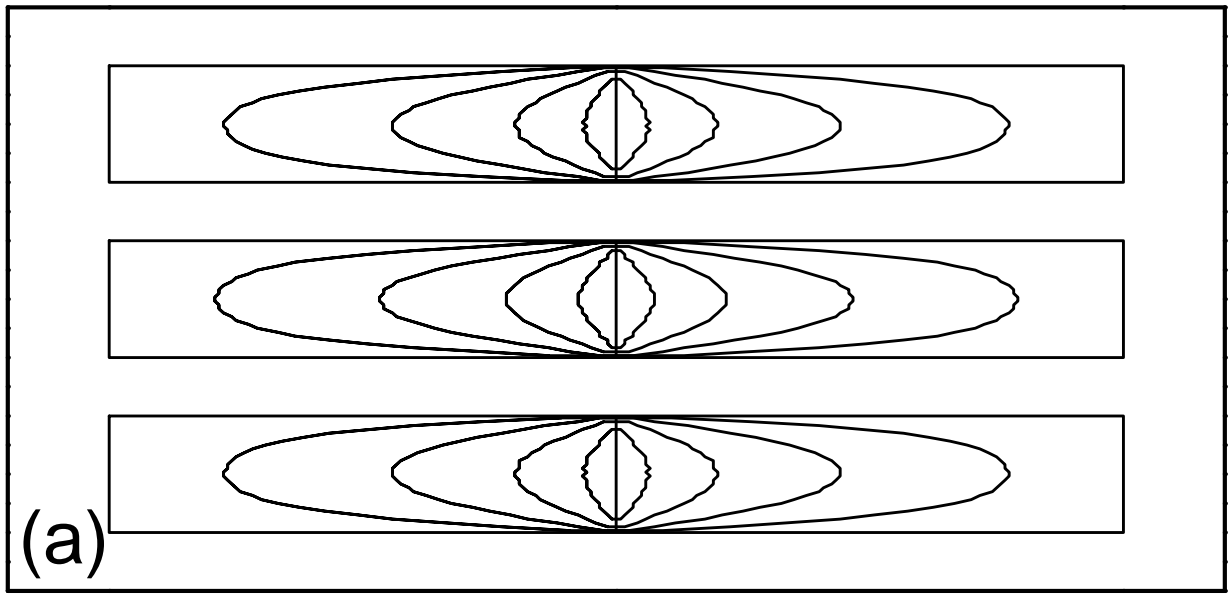
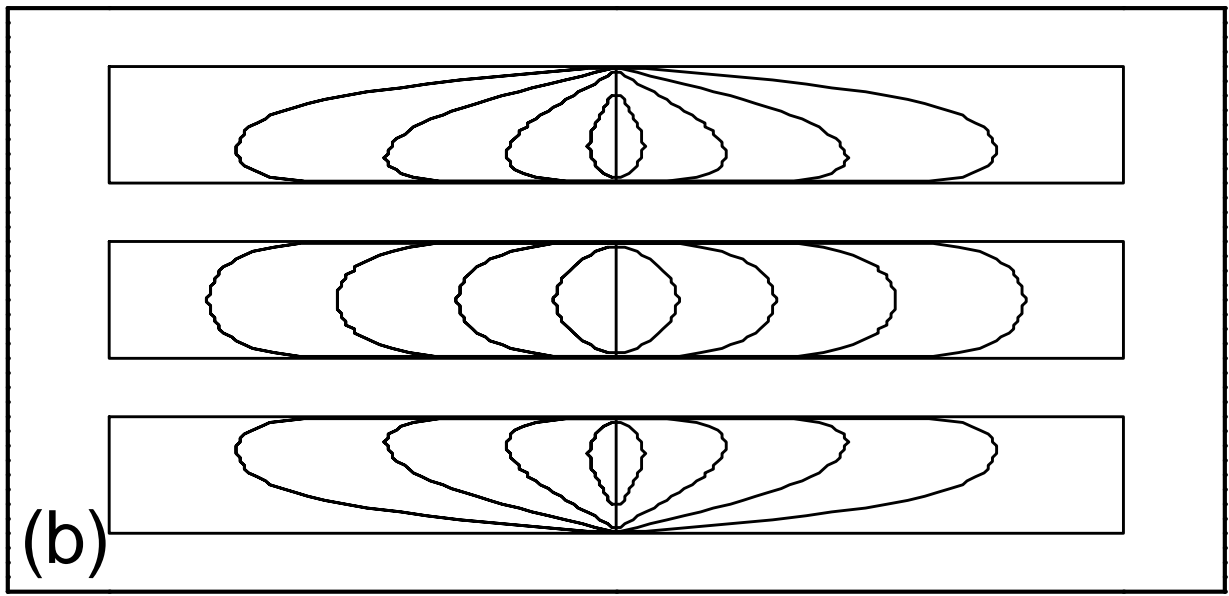
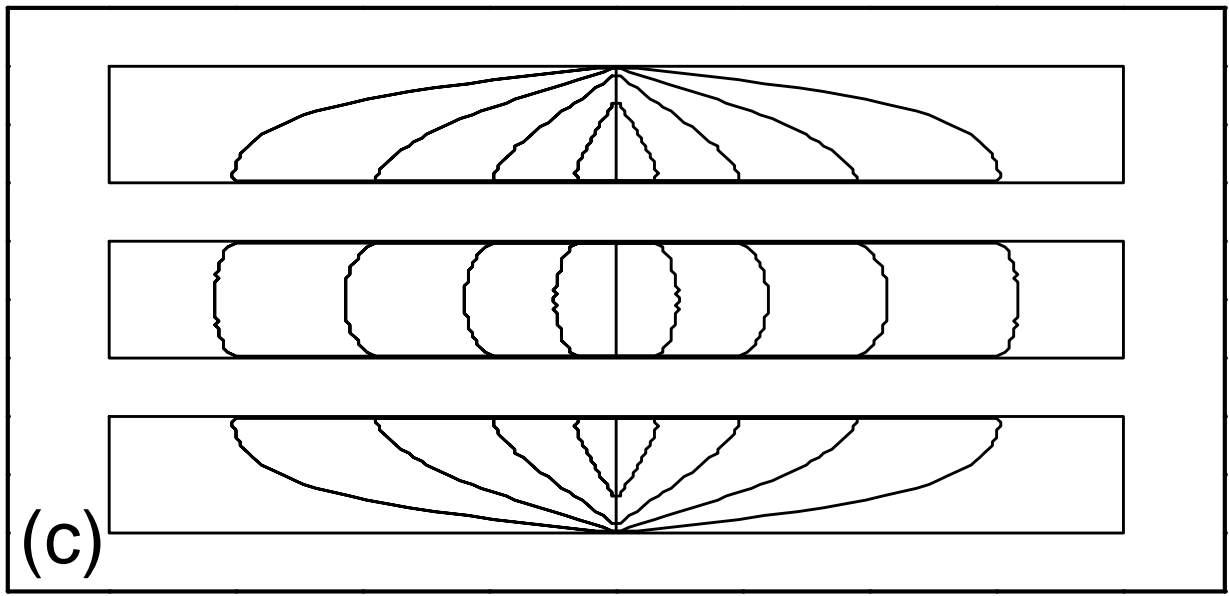
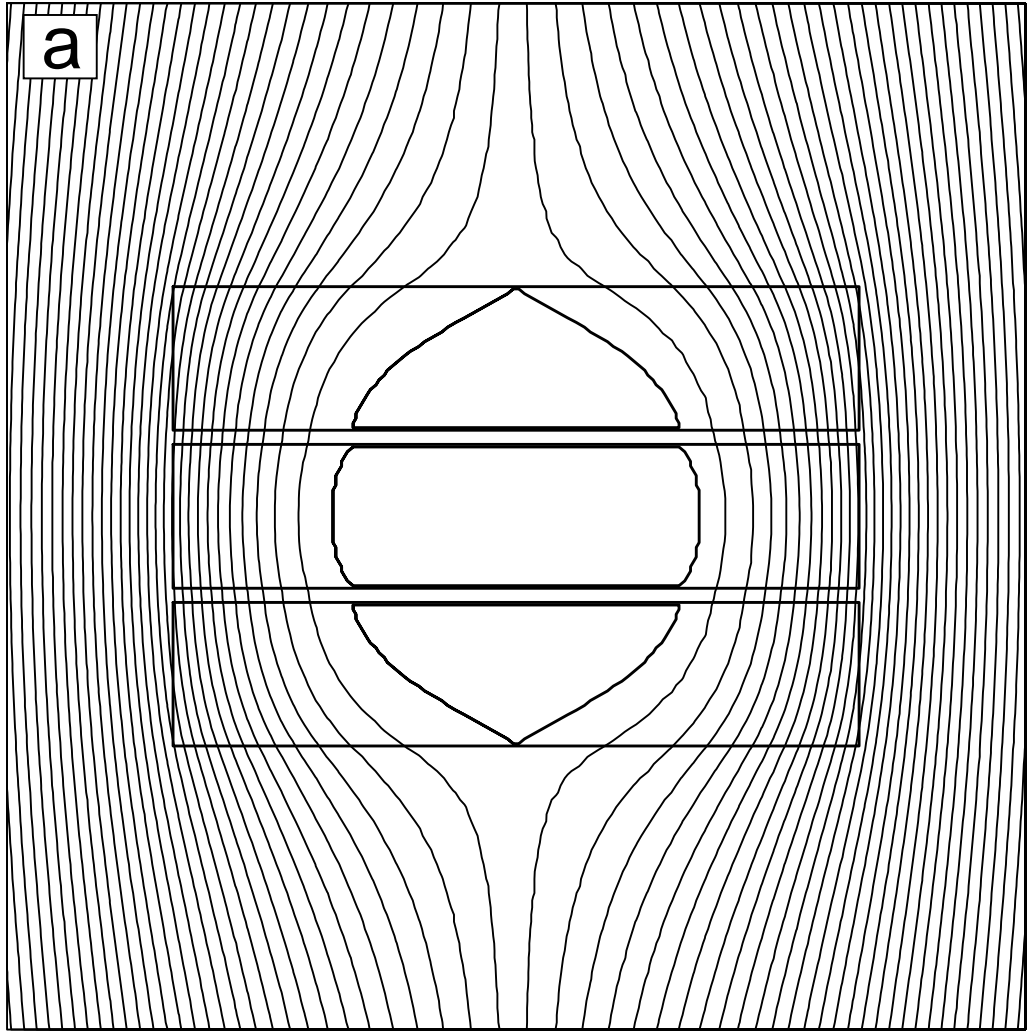
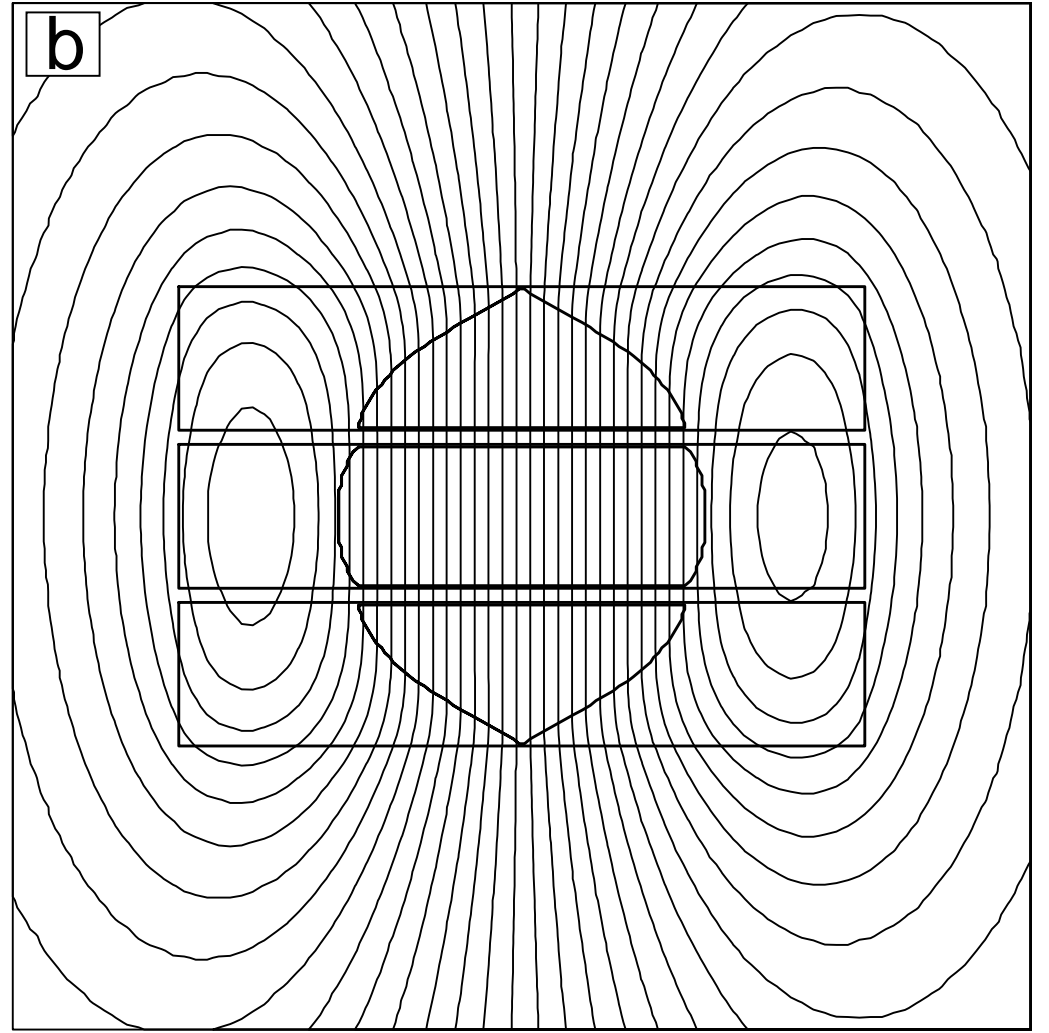
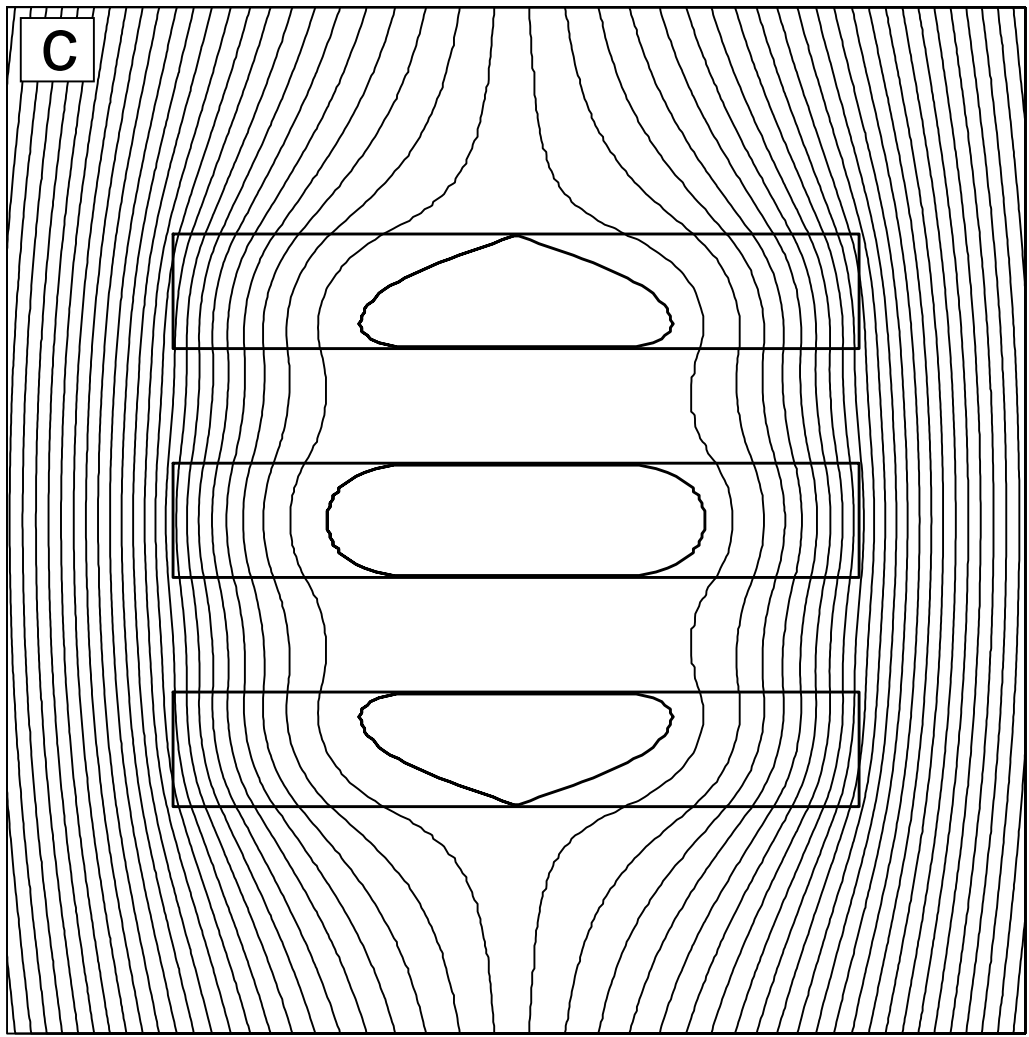
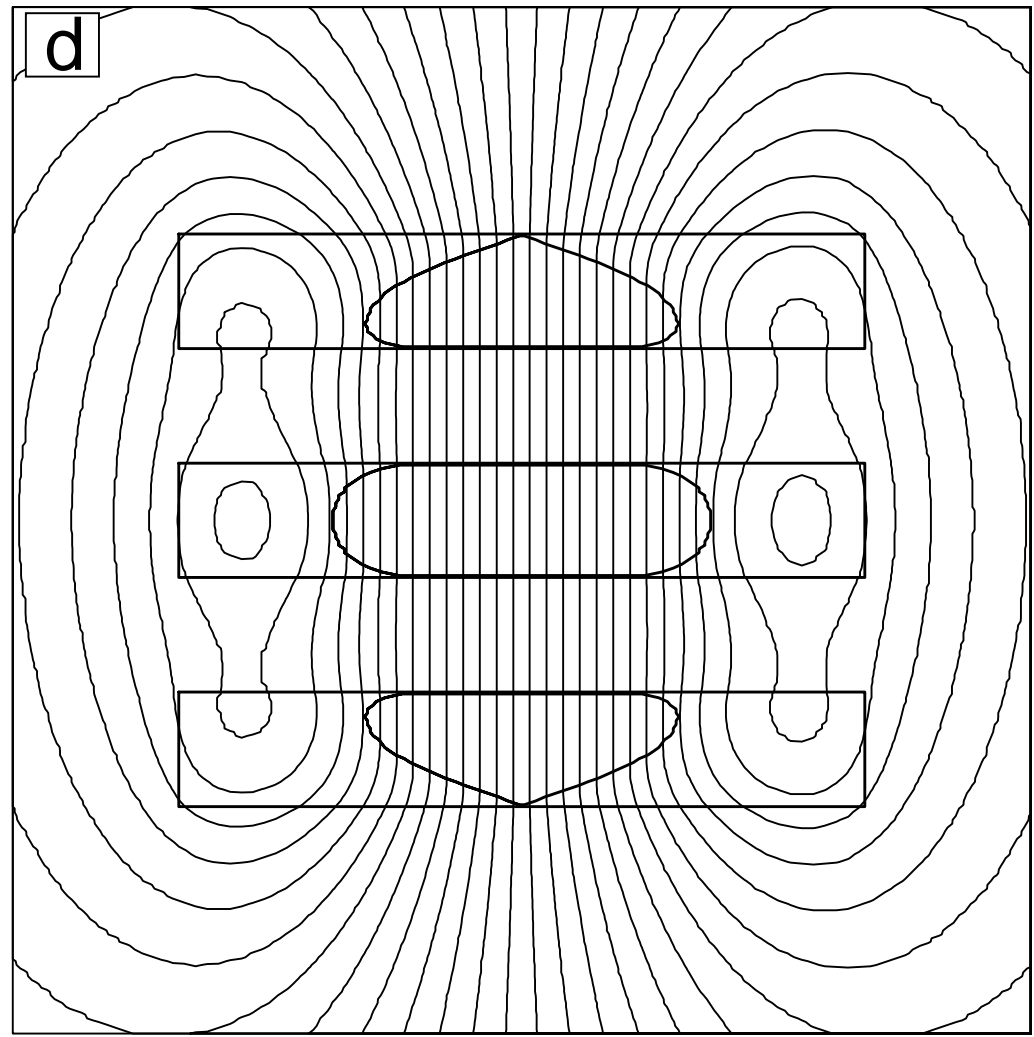
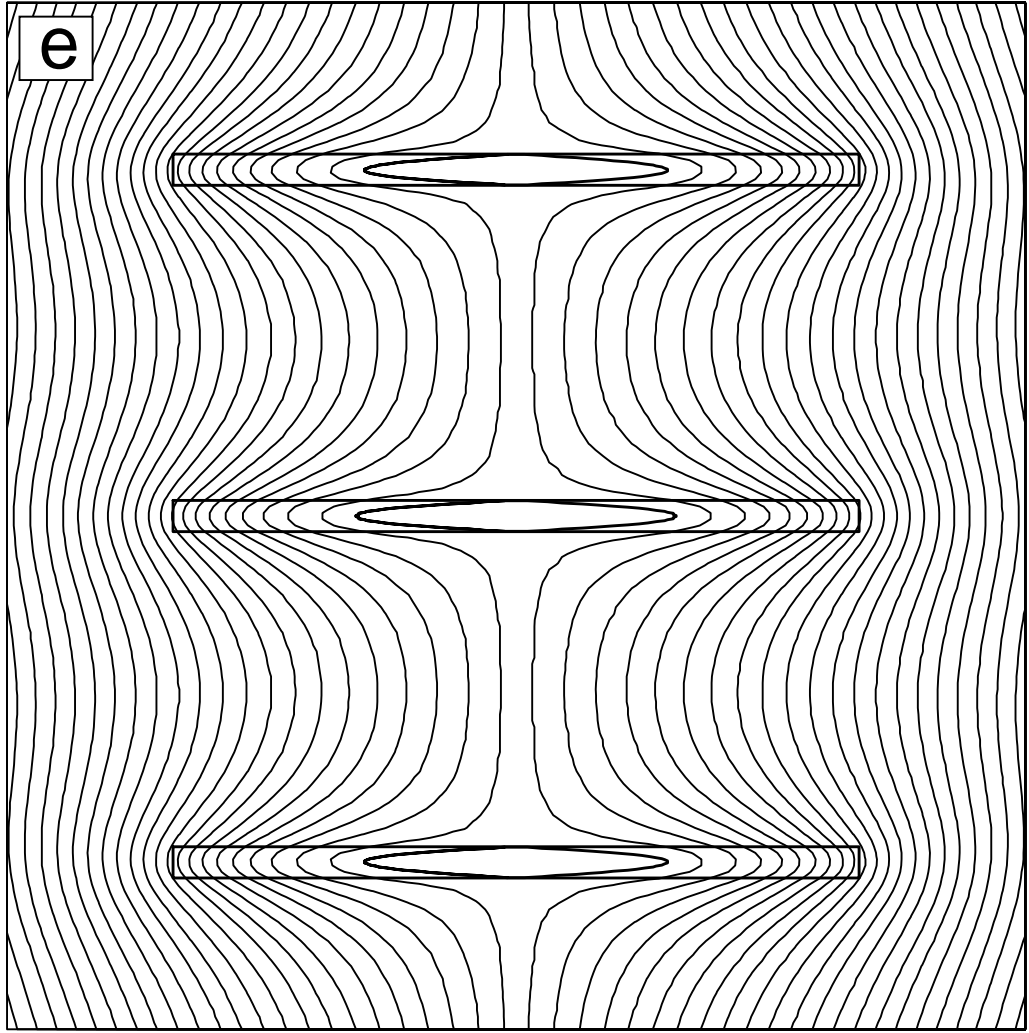
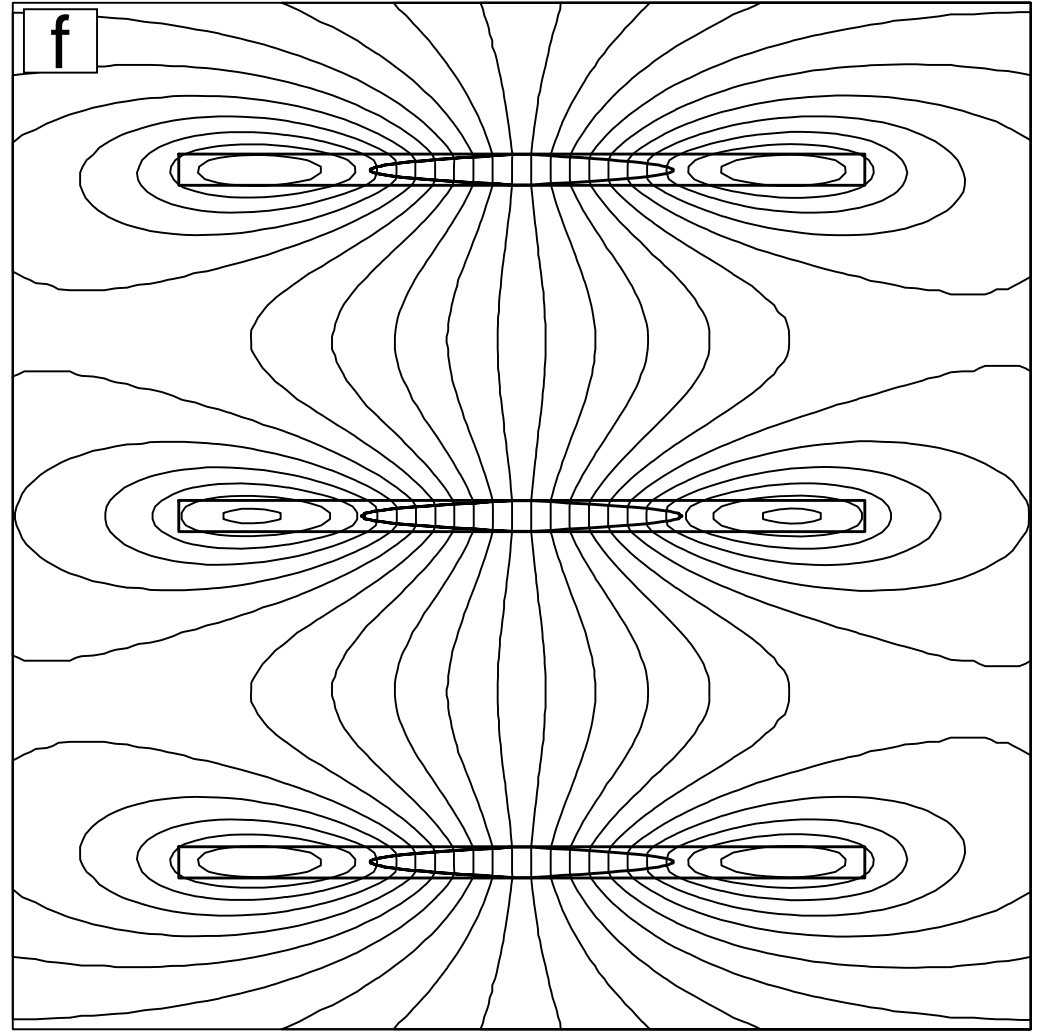
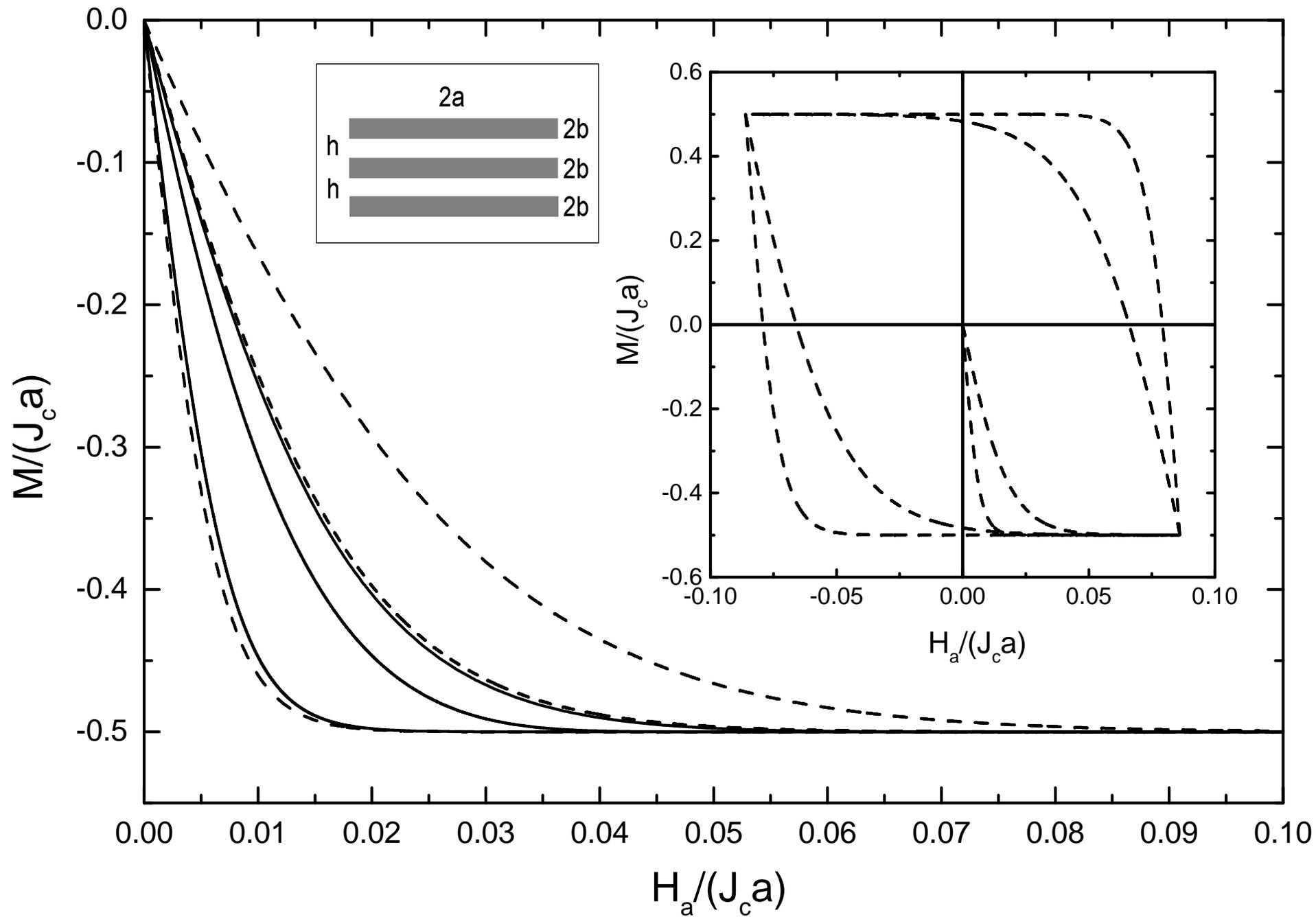
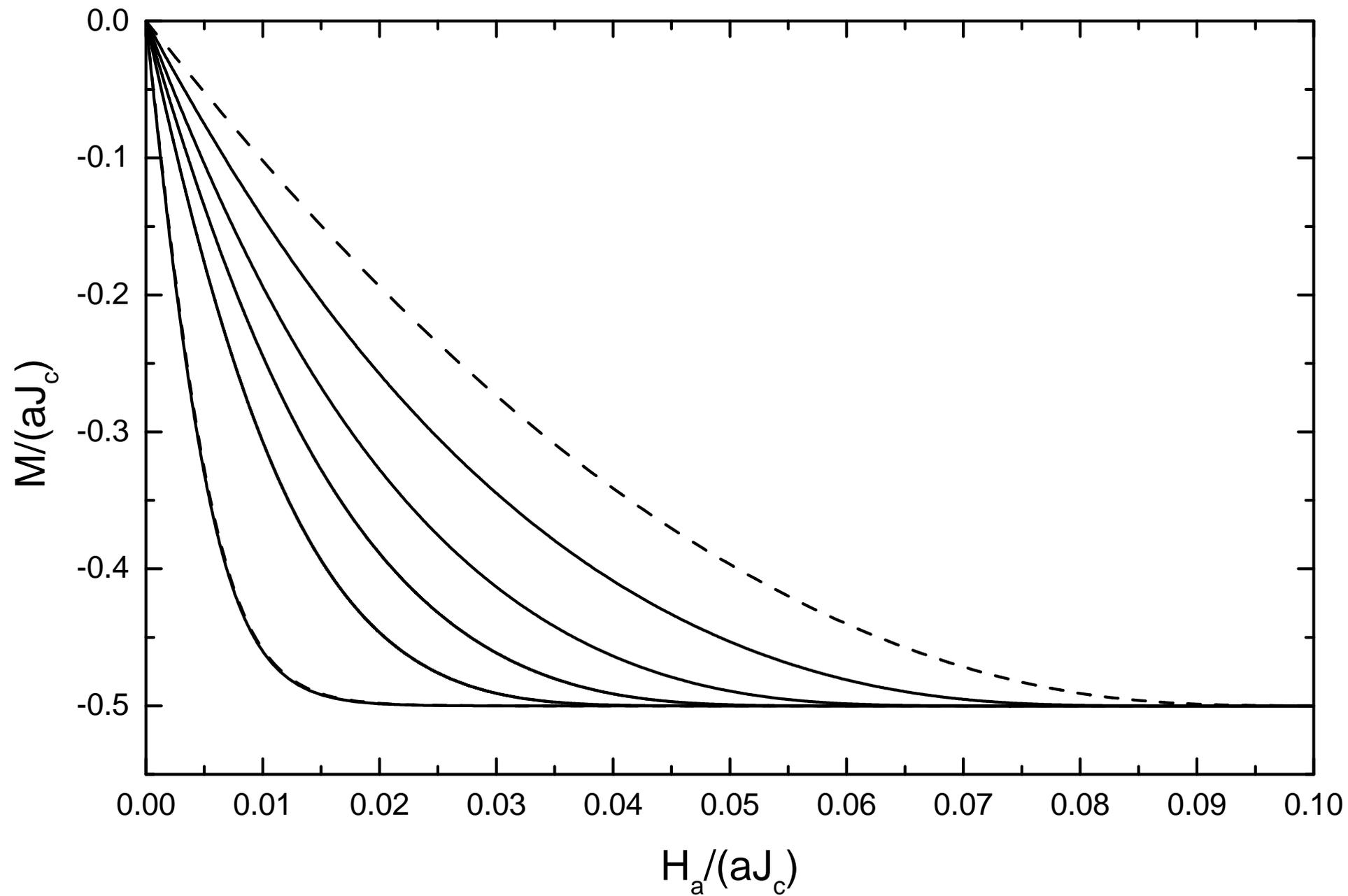
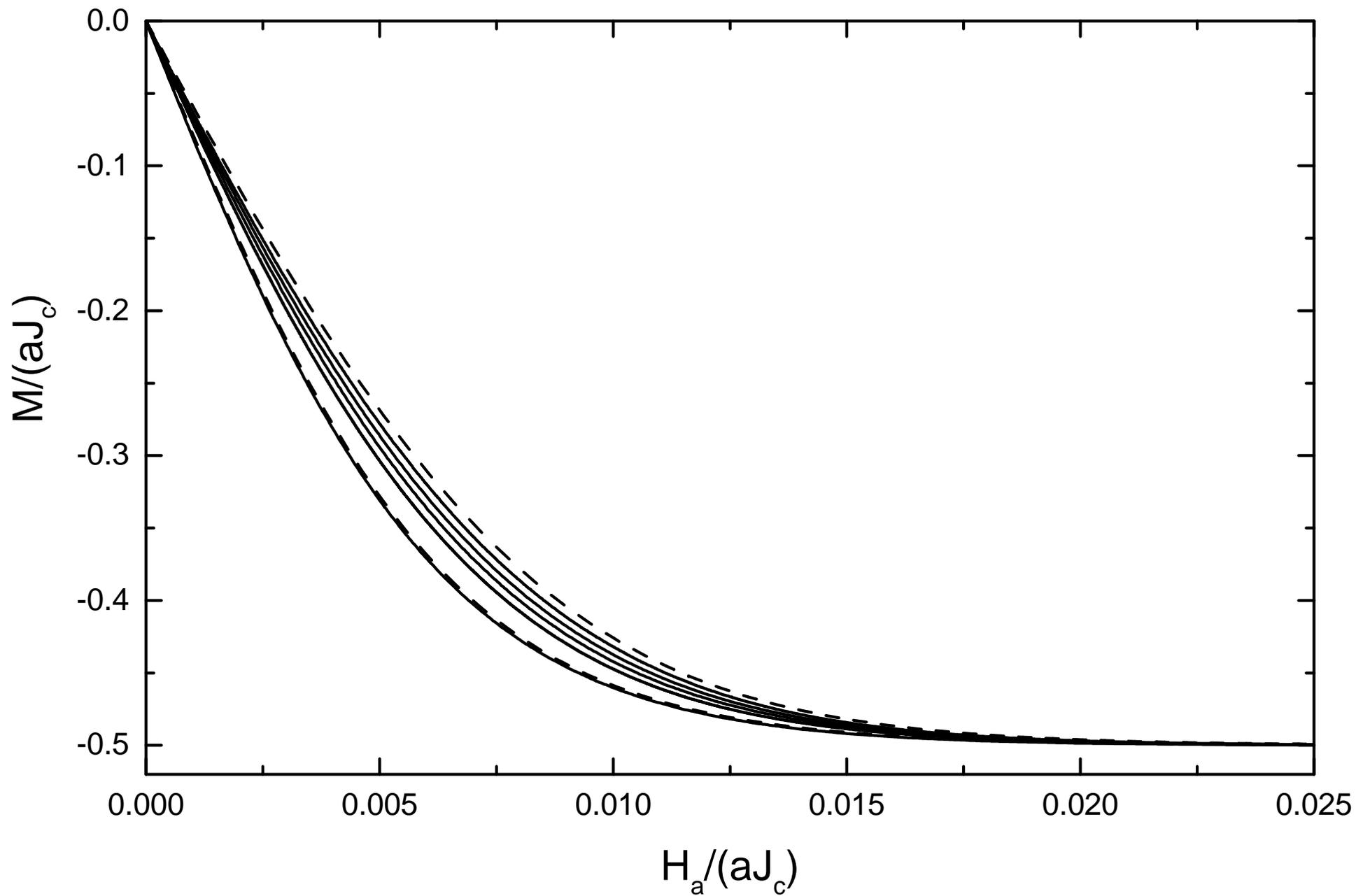
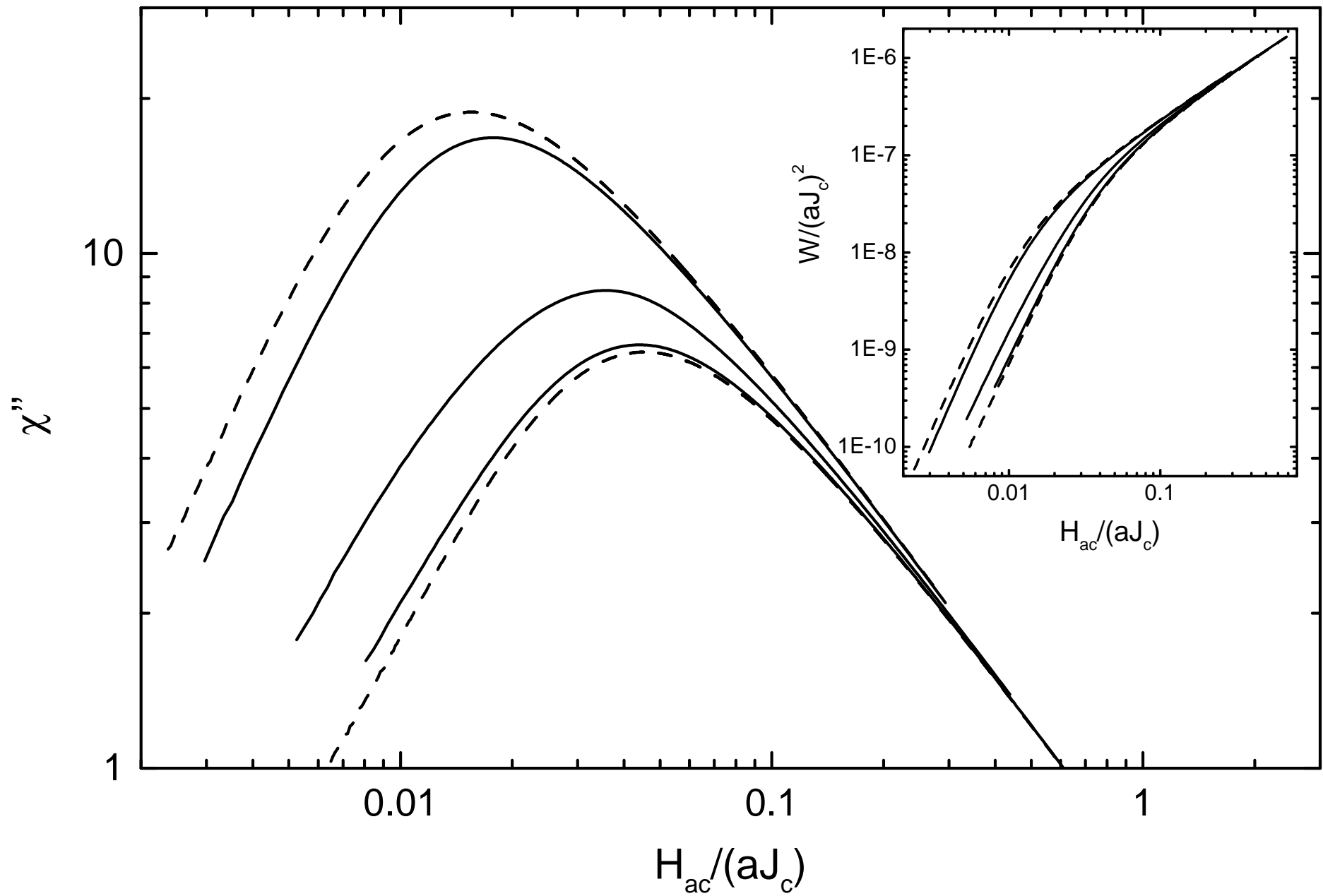
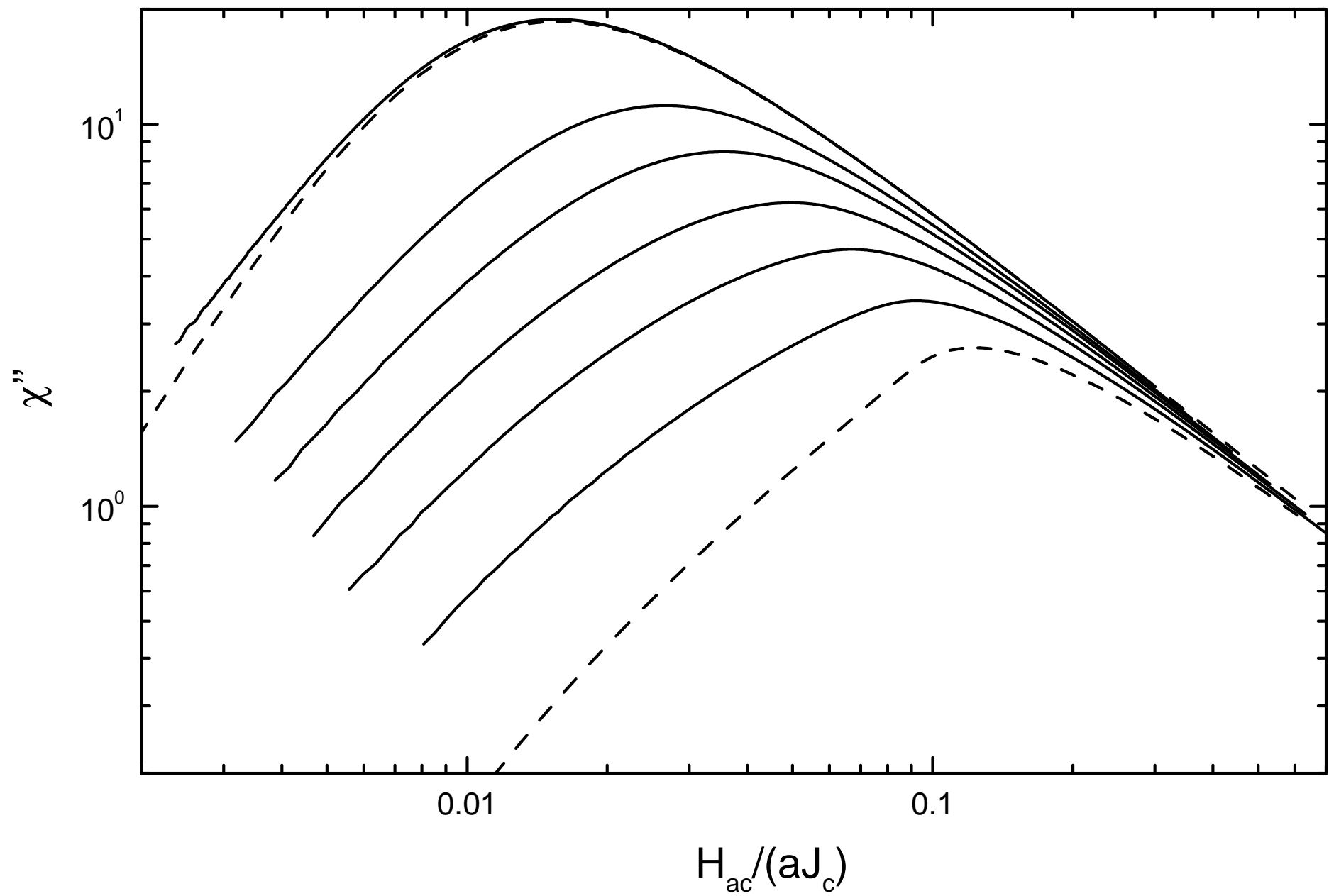
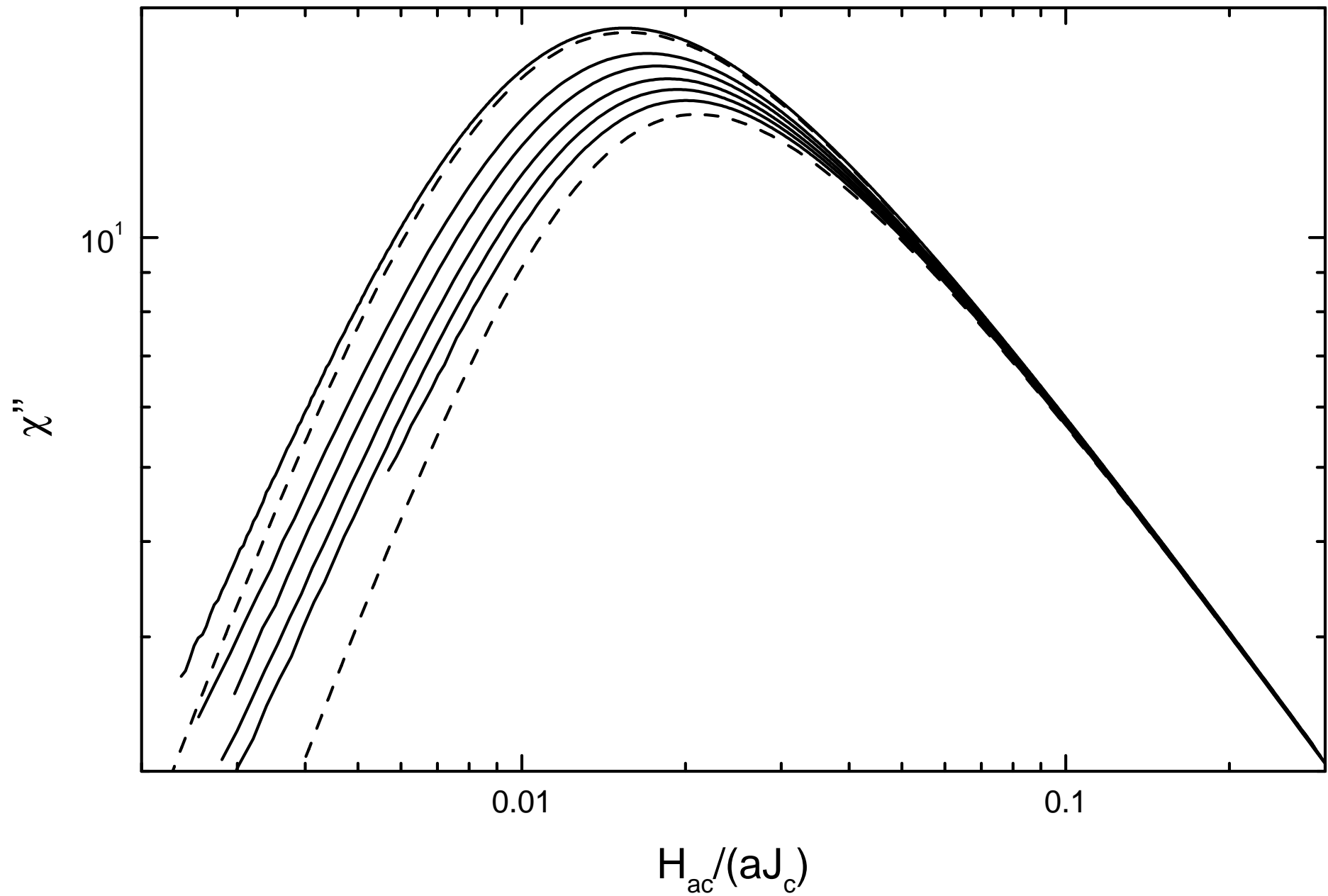
\end{document}